# Gate tunable WSe$_2$/SnSe$_2$ backward diode with ultra-high reverse rectification ratio

Krishna Murali[||], Medha Dandu[||], Sarthak Das, and Kausik Majumdar[*]

Department of Electrical Communication Engineering, Indian Institute of Science, Bangalore 560012, India

[||]These authors contributed equally

[*]Corresponding author, email: kausikm@iisc.ac.in

**ABSTRACT: Backward diodes conduct more efficiently in the reverse bias than in the forward bias, providing superior high frequency response, temperature stability, radiation hardness, and 1/f noise performance than a conventional diode conducting in the forward direction. Here we demonstrate a van der Waals material based backward diode by exploiting the giant staggered band offsets of WSe$_2$/SnSe$_2$ vertical heterojunction. The diode exhibits an ultra-high reverse rectification ratio ($R$) of $\sim 2.1 \times 10^4$ and the same is maintained up to an unusually large bias of $1.5$ V – outperforming existing backward diode reports using conventional bulk semiconductors as well as one- and two-dimensional materials by more than an order of magnitude, while maintaining an impressive curvature coefficient ($\gamma$) of $\sim 37$ V$^{-1}$. The transport mechanism in the diode is shown to be efficiently tunable by external gate and drain bias, as well as by the thickness of the WSe$_2$ layer and the type of metal contacts used. These results pave the way for practical electronic circuit applications using two-dimensional materials and their heterojunctions.**

*Introduction:*

Heterostructures formed by incorporating different semiconductor materials play a vital role in the modern semiconductor world as they offer unique properties employing complementary features of each material. The distinctive electronic and optoelectronic characteristics of heterojunctions have been deployed in a wide variety of solid-state devices[1]. However, lattice matching across the interface is critical in achieving desired performance as any epitaxial distortion induces detrimental interface defect states. Van der Waals force across layered two-dimensional (2D) materials eases the heterostructure fabrication from this aspect[2]. Vertical stacking of these 2D materials opens the possibility of assembling different layers arbitrarily without any deliberation over the precision of lattice matching unlike their bulk counterparts[3–5]. Recent developments in vapor phase growth techniques facilitate the direct synthesis methods of 2D/2D vertical heterojunctions[6–8]. These 2D heterostructures have been recently investigated for different device applications such as transistors[9–12], photodetectors[13,14], photovoltaics[15,16] and LEDs[17,18].

Backward diode[1,19–21] is a special type of diode that conducts more efficiently in the reverse direction than in the forward direction, typically due to strong Zener tunneling under reverse bias. These diodes do not suffer from minority carrier storage induced capacitance as in a diffusion controlled forward biased conventional diode operation. Consequently, backward diodes find widespread applications in high frequency switches, microwave detectors and mixers[22–25]. The low threshold conduction in the reverse bias is also useful for small peak to peak signal rectification applications. Backward diodes typically provide superior temperature stability, radiation hardness, and $1/f$ noise performance compared to conventional diodes[26,27].

SnSe$_2$, a tin based transition metal dichalcogenide, exhibits a large electron affinity[28] with degenerate n-type doping. On the other hand, WSe$_2$ shows ambipolar behavior[29] where the characteristics can be modulated by gate voltage and contact metal. The band offset between SnSe$_2$ and WSe$_2$ forces a highly staggered type II band alignment[30–32]. In this work, we exploit this property of the WSe$_2$/SnSe$_2$ heterojunction to demonstrate a backward diode with a large curvature coefficient of $\sim 37$ V$^{-1}$, coupled with an extremely high reverse rectification ratio of $\sim 2.1 \times 10^4$, outperforming previously reported numbers[19,20,33–43]. We also demonstrate an efficient modulation of the rectification ratio by tuning the applied gate voltage, contact metals, and thickness of the WSe$_2$ layer. Finally, we show that the effective current transfer length at the heterointerface of these vertical heterojunctions can be very large encompassing the entire overlap area of WSe$_2$ and SnSe$_2$, avoiding current crowding, which is in sharp contrast to typical metal-2D semiconductor contact interfaces, with small transfer length[44,45].

**Results and Discussions:**

We first study the degree of charge transfer across WSe$_2$/SnSe$_2$ heterojunction owing to the highly staggered type II band alignment (Figure 1a)[46,47]. We employ photoluminescence (PL) spectroscopy to obtain first-hand information about charge transport processes across the interface[44]. To have an easy optical access to the photoactive WSe$_2$ layer, for the PL experiment, we make a WSe$_2$-top/SnSe$_2$-bottom heterostructure (see **Experimental Section**) on a Si wafer coated with 285 nm SiO$_2$, as schematically depicted in Figure 1b. An optical image of the structure is shown in the top panel of Figure 1c. Atomic Force Microscopy (AFM) along the white dashed line in Figure 1c suggests that the thickness of WSe$_2$ and SnSe$_2$ flakes are 4.7 nm and 22.6 nm, respectively (Figure 1d). The Raman spectra of different portions of the flakes, as characterized with a 532-nm laser illumination, are shown in Figure 1e. The isolated WSe₂ portion exhibits strong $A_{1g}$ and 2LA(M) peaks at 251 cm⁻¹ and 258 cm⁻¹ respectively [48], which are suppressed at the junction area. On the other hand, the junction clearly shows the $E_g$ and $A_{1g}$ peaks of SnSe₂ at 110 cm⁻¹ and 184.5 cm⁻¹ respectively.

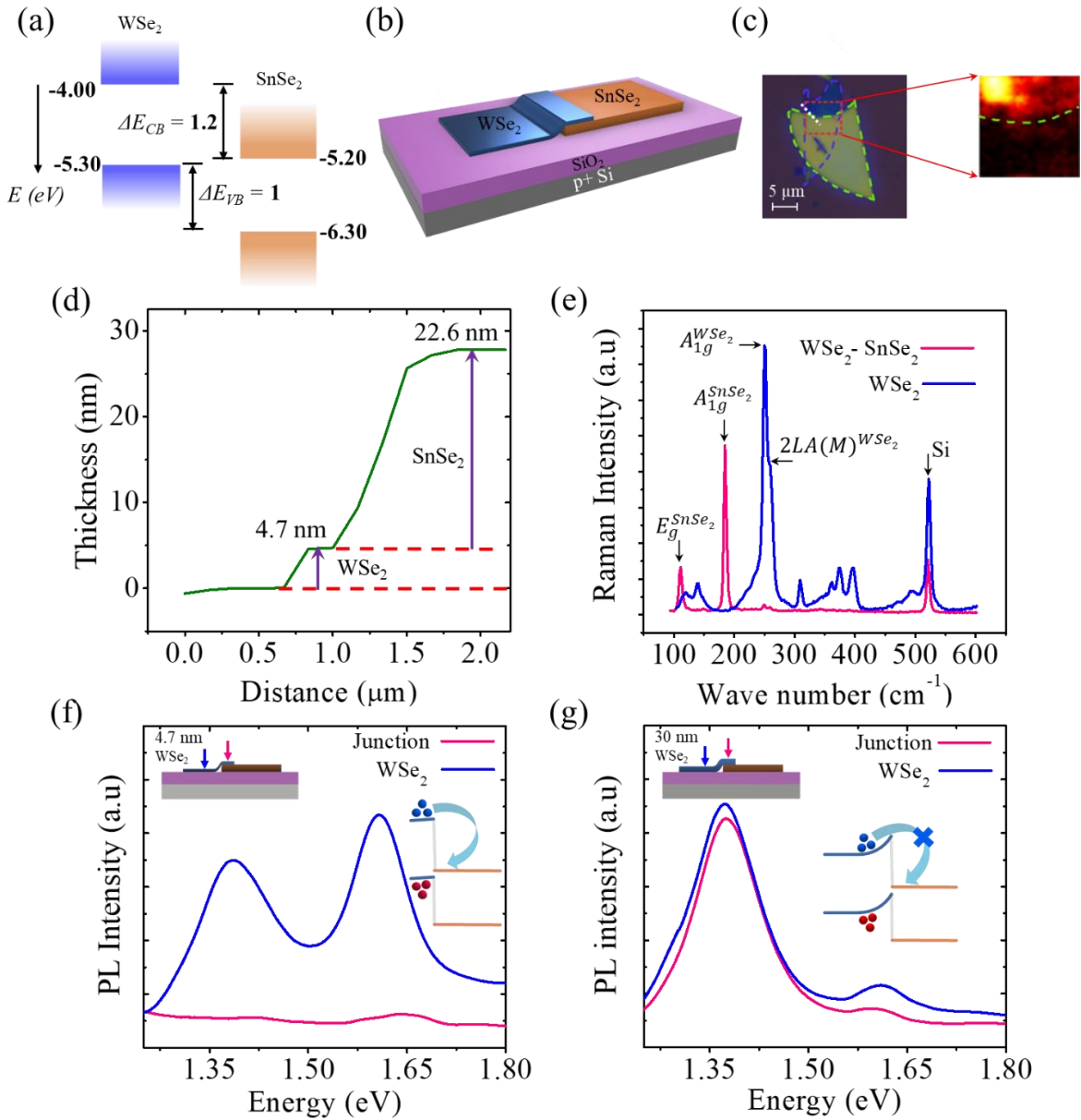

**Figure 1:** (a) Band alignments of bulk $WSe_2$ and $SnSe_2$. (b) Schematic of $WSe_2$-top/$SnSe_2$-bottom heterojunction used for PL mapping. (c) Optical image of the heterojunction used for the PL experiment and the corresponding PL map of the region surrounded by red dashed lines. The dark portion from the junction suggest strong PL quenching. (d) Thickness characterization from AFM depicting thickness profiles of $WSe_2$ (4.7 nm) and $SnSe_2$ (22.6 nm) regions along the white dotted line in (c). (e) Raman spectra (using 532 nm laser) measured on isolated $WSe_2$ portion and on $WSe_2/SnSe_2$ junction. (f-g) PL spectra of (f) thin (~4.7 nm) and (g) thick (~30 nm) $WSe_2$ flake, in isolated (blue) and in junction (pink) area. Top Inset, Schematic of the cross section of the structure. Right inset, Band diagram illustrating charge transfer. Band bending in $SnSe_2$ is negligible owing to its degenerate doping.

In Figure 1c, we show the $WSe_2$ indirect peak PL intensity mapping of the area surrounded by the red dashed lines. We observe almost complete quenching of PL intensity at the junction, compared with the isolated $WSe_2$ portion. Figure 1f shows the corresponding PL spectra of the characterized junction and isolated $WSe_2$ portion. The large conduction band offset of ~1.2 eV at the $WSe_2/SnSe_2$ junction as described in Figure 1a suggests efficient transfer of the photo-excited electrons from $WSe_2$ to $SnSe_2$, supporting the $WSe_2$ PL quenching. Interestingly, the PL quenching is not significant when the thickness of the $WSe_2$ flake is increased. For example, the PL spectra for a 30-nm thick $WSe_2$ flake on $SnSe_2$ is shown in Figure 1g. Such thickness dependent PL quenching can be explained from the band bending on the $WSe_2$ side in both the investigated structures. Under equilibrium, to maintain the band offsets at the $WSe_2/SnSe_2$ interface and zero field at the top $WSe_2$/air interface, the bands of $WSe_2$ in the depletion region are forced to bend concave downward from $WSe_2/SnSe_2$ interface, which is supported by the moderate p-doping in $WSe_2$. When $WSe_2$ flake is too thin, bands cannot bend steeply in the constrained physical space to reach equilibrium positions as illustrated in the inset of Figure 1f. Here, the photo-excited electrons in $WSe_2$ get transferred to $SnSe_2$ side because the minimal band bending in $WSe_2$ does not form any barrier. But in the case of thick $WSe_2$ as in the inset of Figure 1g, bands can bend sufficiently, which forms a barrier for the electrons inhibiting their transfer to $SnSe_2$.

Before discussing the electrical transport properties of WSe$_2$/SnSe$_2$ junction, we discuss the individual behaviour of SnSe$_2$ and WSe$_2$ channels using back gated structures, as summarized in Figure 2. SnSe$_2$ is a degenerately n-doped semiconductor with Fermi-level lying about 0.2-0.3 eV above the conduction band[49]. We obtained similar values using KPFM measurements[50]. This results in ohmic and highly conductive $I_{ds}$-$V_{ds}$ response from SnSe$_2$ channel with Ni/Au contacts, as shown in Figure 2a. The degenerate doping results in negligible modulation of current by the back gate voltgae ($V_g$).

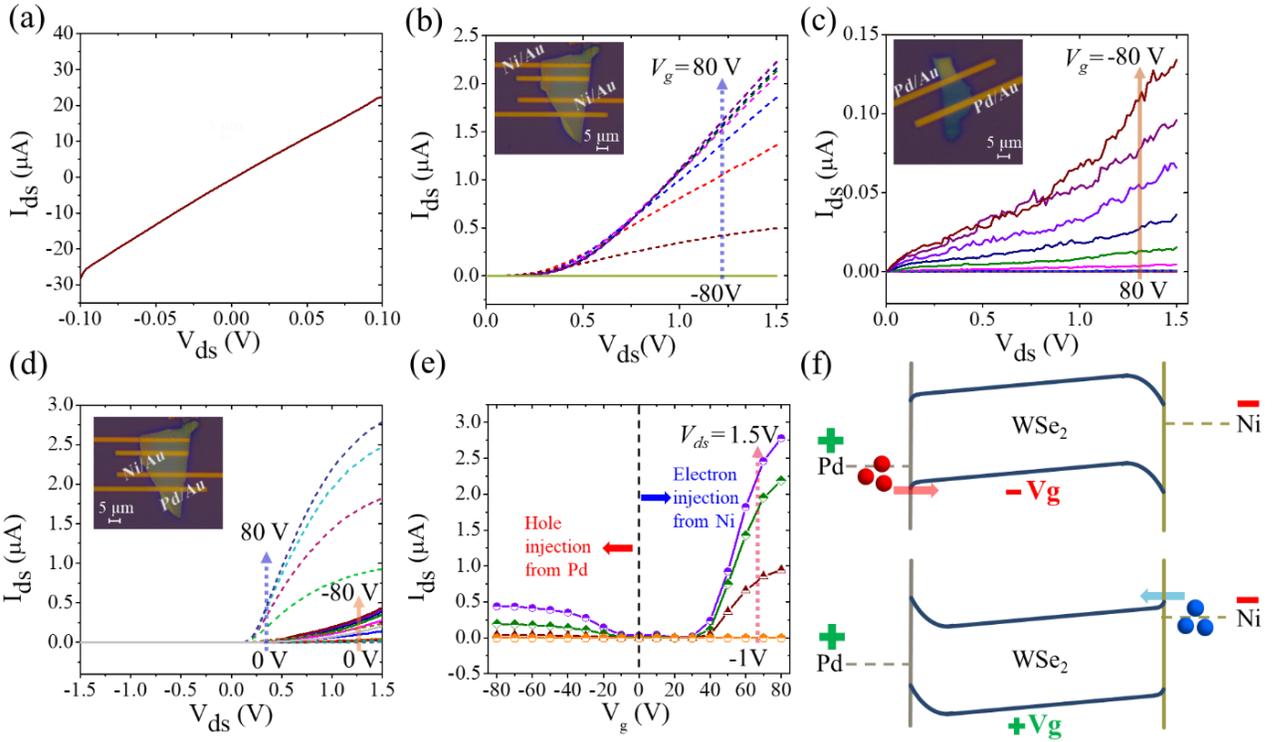

**Figure 2**: (a) $I_{ds}$-$V_{ds}$ characteristics across the two Ni/Au contacts on SnSe$_2$. (b-c) $I_{ds}$-$V_{ds}$ characteristics of back gated WSe$_2$ channel with (b) two Ni/Au contacts and (c) two Pd/Au contacts. $V_g$ varies from −80 V to 80 V in steps of 10 V. Inset, an optical image of the fabricated device. (d-e) $I_{ds}$-$V_{ds}$ and $I_{ds}$-$V_g$ characteristics of back gated asymmetric WSe$_2$ channel with Pd/Au and Ni/Au contacts. Ni/Au is grounded while bias is applied to Pd/Au. Inset of (d), an optical image of the fabricated device. (f) Band diagrams of (d) and (e), under forward bias, with negative (top panel) and positive (bottom panel) gating. Under negative $V_g$, Pd efficiently injects holes (red

spheres) into the WSe$_2$ channel, while under positive $V_g$, Ni injects efficiently electrons (blue spheres).

On the other hand, as explained in Figures 2b-c, Ni contacted WSe$_2$ behaves as an nFET with an increasing current for $V_g > 0$, while Pd contacted WSe$_2$ channel behaves like a pFET with an increasing current for $V_g < 0$. These observations suggest that Ni (Pd) can efficiently inject electrons (holes) into the WSe$_2$ channel. Consequently, the asymmetric device with Ni/Au and Pd/Au as two different contacts to WSe$_2$ channel exhibits a perfectly rectifying, but ambipolar behavior, as shown in Figures 2d-e. Figure 2f explains the forward bias current mechanism under both gating conditions. At positive gate voltage, under forward bias (Ni side grounded, Pd side biased), Ni efficiently injects electrons, providing large current. However, under reverse bias condition, Pd being poor injector of electrons into WSe$_2$, the current is suppressed. On application of negative gate voltage, under forward bias condition, Pd efficiently injects holes. However, poor hole injection by Ni under reverse bias condition suppresses the current.

We next fabricate a back-gated heterojunction diode D1 using the WSe$_2$/SnSe$_2$ stack as illustrated in Figure 3a with Ni/Au contact on SnSe$_2$ and Pd/Au contact on WSe$_2$ (see **Experimental Section**). The thickness of the WSe$_2$ and the SnSe$_2$ flakes are 8 nm and 122 nm, respectively, as suggested by AFM characterization in Figure 3b. Contrary to the PL experiment structure, we now use WSe$_2$-bottom/SnSe$_2$-top heterostructure to efficiently modulate ambipolar WSe$_2$ by the back gate. The $V_g$ dependent current-voltage characteristics, as summarized in Figure 3c-d, show three important features. First, the magnitude of the current increases with both negative and positive $V_g$, retaining the ambipolar behavior of WSe$_2$. Second, both the forward ($I_{forward}$) and reverse ($I_{reverse}$) bias currents can be efficiently controlled by 6 orders of magnitude by tuning $V_g$. Third, the diode conducts heavily in the reverse bias regime than in the forward bias, particularly for $V_g > 0$, and hence can be used as a backward diode. In the inset of Figure 3d, we show the turn on behavior of the diode in the linear scale under reverse bias, with suppressed conduction in the forward bias. The reverse rectification ratio ($R = \frac{I_{reverse}}{I_{forward}}$) is plotted as a function of $V_g$ and $|V_{ds}|$ in Figure 3e. Here $I_{reverse}$ and $I_{forward}$ are measured at $-V_{ds}$ and $+V_{ds}$, respectively. $R$ is found to reach an impressive value of ~220 for small positive $V_g$ and large $|V_{ds}|$.

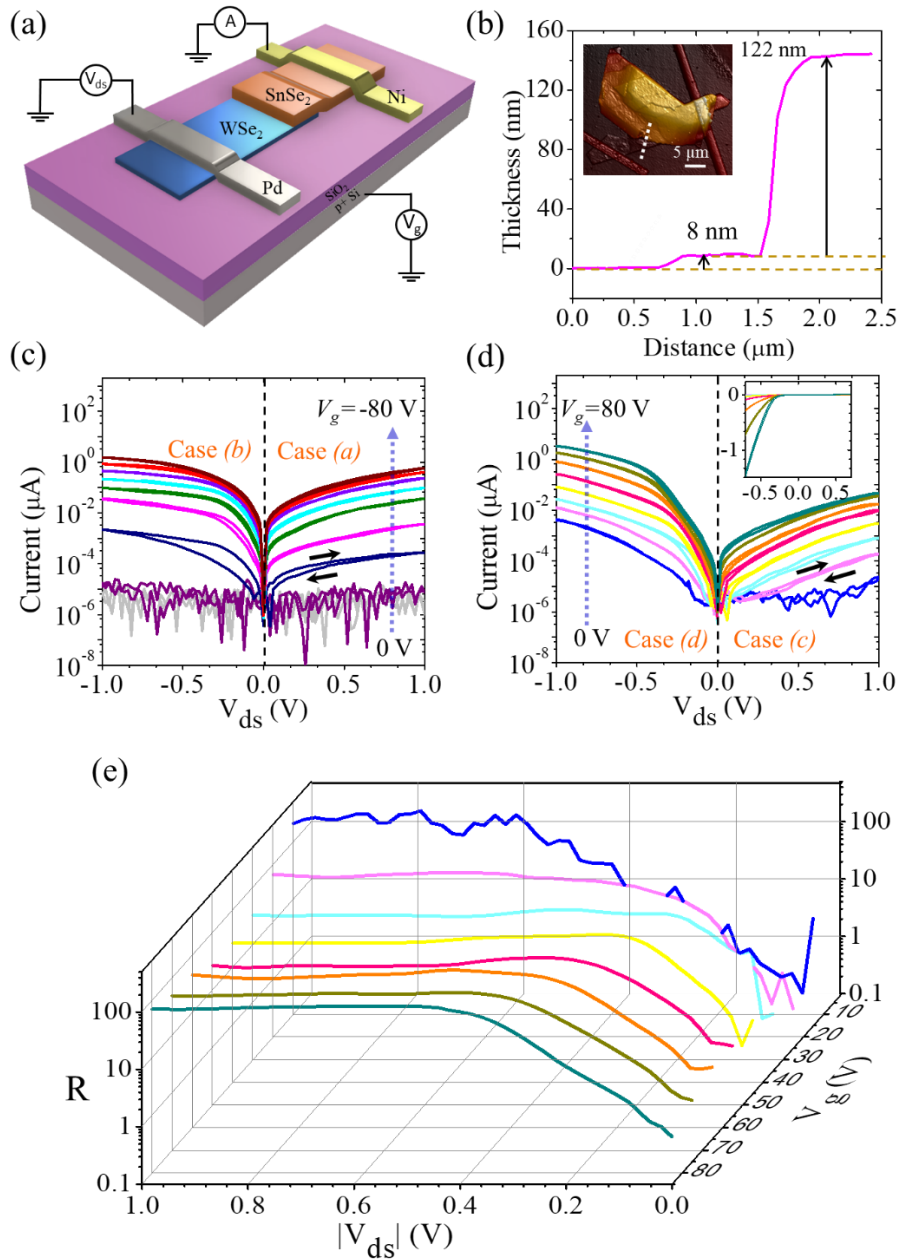

**Figure 3:** (a) Schematic of back gated WSe₂-bottom/SnSe₂-top device (D1) fabricated for electrical measurement. (b) Thickness profile of flakes along the white dashed line on AFM image of the device in the inset. (c) Log scale $I_{ds}$-$V_{ds}$ characteristics of the device under negative gating, starting from 0 V to −80 V in steps of 10 V. The black arrows indicate the forward and reverse sweep, with negligible hysteresis. (d) Log scale $I_{ds}$-$V_{ds}$ characteristics of the device under negative gating, starting from 10 V to 80 V in steps of 10 V. Inset, Linear scale plot of $I_{ds}$-$V_{ds}$. (e) Reverse rectification ratio ($R = \frac{I_{reverse}}{I_{forward}}$) of D1 as a function of drain bias and positive gate voltage. $I_{reverse}$ and $I_{forward}$ are measured at −$V_{ds}$ and +$V_{ds}$, respectively.

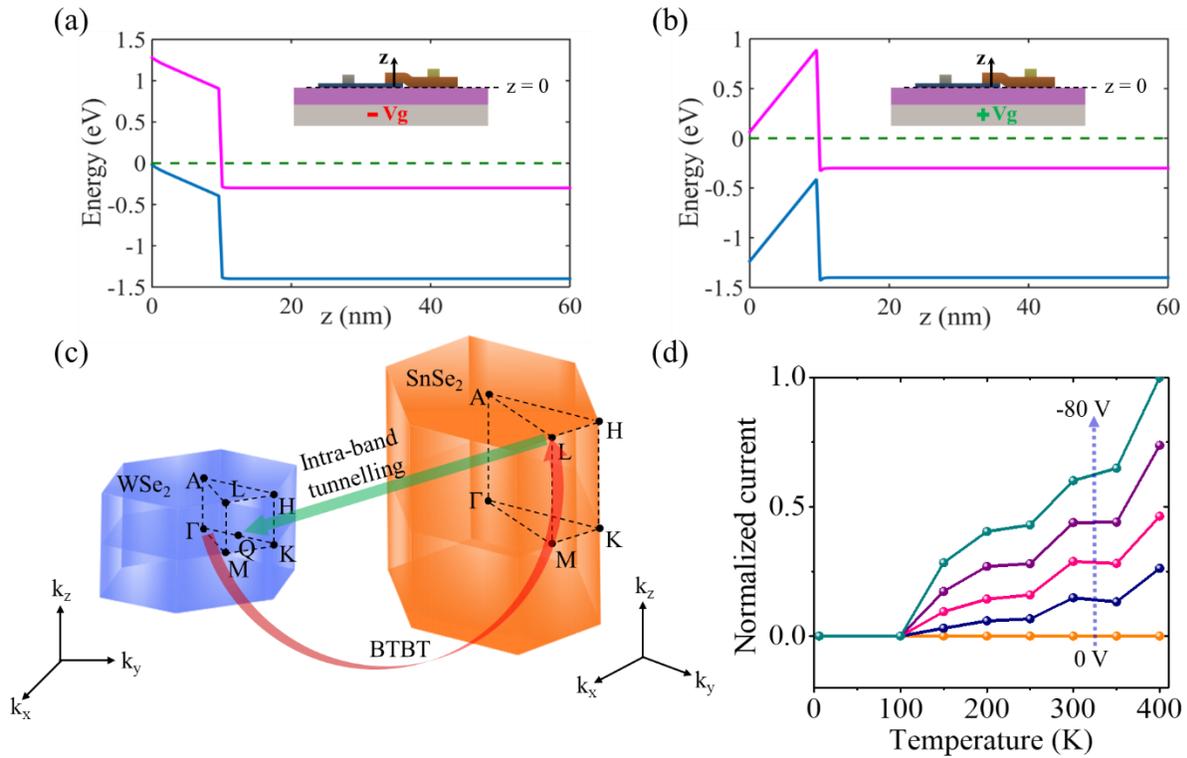

**Figure 4:** (a)-(b) Simulated 1D equilibrium band diagrams across the vertical cross-section of the heterojunction under (a) negative and (b) positive gating. The green dashed line indicates the chemical potential. (c) First Brillouin zones of bulk WSe₂ (blue) and bulk SnSe₂ (orange). Band to Band tunnelling (BTBT) across VB of WSe₂ and CB of SnSe₂ is shown by red arrow while green arrow depicts the intra-band electron tunnelling from CB of SnSe₂ to CB of WSe₂. (d) Normalized current across WSe₂/SnSe₂ device at $V_{ds} = -0.5$ V, as a function of temperature at gate voltages varying from 0 V to −80 V in steps of 10 V.

To understand the large magnitude of reverse current from the nature of the WSe₂/SnSe₂ vertical heterojunction, we self-consistently solve 1D Poisson equation in conjunction with semi-classical charge density to obtain the band bending along the vertical $z$ direction, with the WSe₂/oxide interface being taken as $z = 0$. The resulting band diagrams in equilibrium are shown in Figure 4a-b for negative and positive $V_g$. The band diagrams clearly indicate that the thickness of the WSe₂ film and the magnitude of $V_g$ play a key role in determining the vertical field in WSe₂, while the SnSe₂ bands remain nearly flat owing to its degenerate doping. From Figure 4a, it is clear that under negative gate voltage, reverse bias would favour strong band-to-band-tunneling (BTBT) of electrons in the vertical direction from valence band of WSe₂ to conduction band of SnSe₂. As shown in Figure 4c, the conduction band minimum (CBM) of SnSe₂ is located at the L point[51] of the Brillouin zone whereas the valence band maximum (VBM) of WSe₂ occurs at the zone center ($\Gamma$ point)[52]. This difference in crystal momentum results in an indirect BTBT of carriers at the heterojunction interface, indicated by the red arrow in Figure 4c. Temperature dependent current measurements at $V_{ds}$ of -0.5 V in Figure 4d shows a strong suppression of the reverse bias tunneling current with reduced temperature, depicting the indirect nature of tunneling.

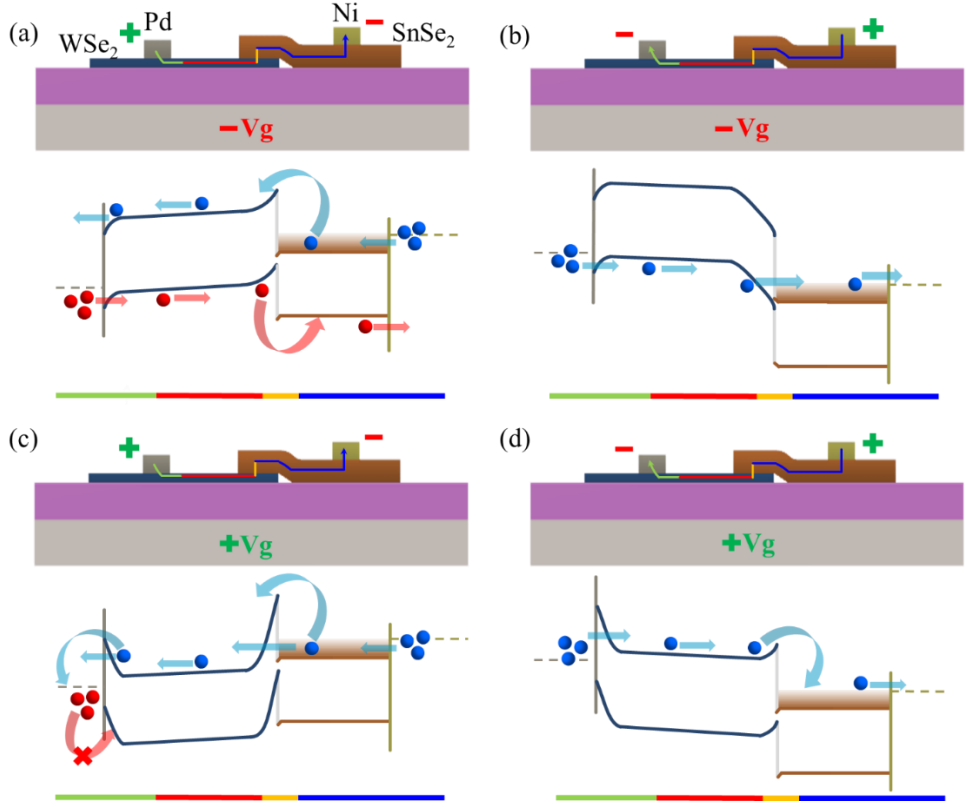

**Figure 5:** (a)-(d) The band diagrams and the corresponding device schematics of diode D1 for the four cases discussed in the text, namely (*a*) $V_g < 0, V_{ds} > 0$, (*b*) $V_g < 0, V_{ds} < 0$, (*c*) $V_g > 0, V_{ds} > 0$, and (*d*) $V_g > 0, V_{ds} < 0$. Color bar below each band diagram and mutli-colored arrow in the respective device cross-section map the relevant regions along which transport is considered.

With the above discussion in background, we next discuss the carrier transport mechanism of Figure 3c-d in four different biasing conditions, namely (*a*) $V_g < 0, V_{ds} > 0$, (*b*) $V_g < 0, V_{ds} < 0$, (*c*) $V_g > 0, V_{ds} > 0$, and (*d*) $V_g > 0, V_{ds} < 0$. The corresponding band diagrams are shown in Figure 5a-d, respectively. The band diagrams are drawn along the arrows, as indicated in the corresponding schematic in each figure, and the different regions are represented by different color coding. In case (*a*), i.e. negative gate voltage and forward bias, Pd contact efficiently injects holes (shown by red spheres) in WSe$_2$, which overcome the WSe$_2$/SnSe$_2$ thermionic hole barrier under forward bias, and eventually get collected at the Ni contact of SnSe$_2$. For a given $V_{ds}$, the hole injection from Pd contact to WSe$_2$ is a strong function of $V_g$, while electron injection from Ni contact to SnSe$_2$ is almost independent of $V_g$. However, the electrons (shown by blue spheres), being injected from Ni contact of SnSe$_2$ have to overcome the SnSe$_2$/WSe$_2$ electron barrier - relatively larger than the hole barrier and get collected by the Pd contact of WSe$_2$. Thus, the holes contribute to a major component of the forward current in case ($a$). For positive $V_g$ and forward bias [case ($c$)], the hole injection from Pd is cut off, suppressing the hole current. On the other hand, positive $V_g$ (Figure 5c) enhances the electron current, as the electrons from the conduction band of SnSe$_2$ can tunnel through to the conduction band of WSe$_2$ through the thin triangular barrier, the width of which is primarily determined by the thickness of the WSe$_2$ film and the magnitude of $V_g$. We note that such intra-band tunneling in the conduction band, also requires a change in crystal momentum, and hence inelastic in nature, as indicated by the green arrow in Figure 4c.

Under reverse bias, with negative $V_g$ [case ($b$)], the current is primarily governed by the indirect BTBT at the WSe$_2$/SnSe$_2$ interface, as discussed earlier. This situation is depicted in Figure 5b, and results in large reverse current. However, under positive $V_g$ [case ($d$)], the mechanism for strong reverse bias current cannot be explained by BTBT, as indicated by Figure 5d. Since the WSe$_2$ film used in the device is ultra-thin (~8 nm), under positive $V_g$, electrons can be injected from the Pd contact to the WSe$_2$ channel by tunneling induced field emission through the Schottky barrier, which eventually are drifted to the Ni contact, with almost zero barrier at larger $V_{ds}$. The WSe$_2$/SnSe$_2$ heterojunction device thus represents an excellent platform where the carrier transport mechanism can be controlled by the external biasing conditions, type of contacts used, and the thickness of the WSe$_2$ film.

To further improve the reverse rectification ratio, we note from Figure 3c-d that the diode works better as a backward diode under positive $V_g$. From the above discussion, $I_{forward}$ under positive $V_g$ is controlled by the intra-band electron tunneling (Figure 5c) through the triangular barrier at the WSe$_2$/SnSe$_2$ interface. Figure 4b suggests that this tunneling barrier is controlled by the thickness of WSe$_2$, and hence one can suppress $I_{forward}$ further by increasing the thickness of WSe$_2$ film. On the other hand, under reverse bias, with positive $V_g$, Ni contact on WSe$_2$ would be a better electron injector than Pd, as explained earlier in Figure 2, owing to Fermi level pinning closer to conduction band edge. Consequently, we next fabricate a WSe$_2$/SnSe$_2$ heterojunction diode D2, as shown in the bottom left inset of Figure 6a, with (i) relatively thick ($\sim 30$ nm) WSe$_2$, (ii) deposit Ni contact both on WSe$_2$ and SnSe$_2$, and (iii) operate the same under positive $V_g$. Figure 6a shows the $I_{ds}$-$V_{ds}$ characteristics in the log scale, suggesting excellent suppression of $I_{forward}$, while $I_{reverse}$ is very strong. The linear scale plot is shown in the top right inset of Figure 6a, indicating near abrupt turn on with low backward threshold voltage ($\sim -0.1$ V). The corresponding $R$ is plotted in Figure 6b as a function of $V_g$ and $|V_{ds}|$. We achieve an ultra-high $R \approx 2.1 \times 10^4$ at $V_g = 0$, which outnumbers the reverse rectification ratio of conventional Si, Ge backward diodes by two orders[24,53]. The characteristics of another device from a different run are shown in Supporting Information S1, showing similar $R$ value. The transport mechanism at forward and reverse bias is explained using the band diagrams in Figure 6c-d. Note that, even though the reverse current in D2 under positive gating is not governed by BTBT, the fact that it is a majority carrier drift driven process, facilitates fast device operation, by avoiding diffusion capacitance.

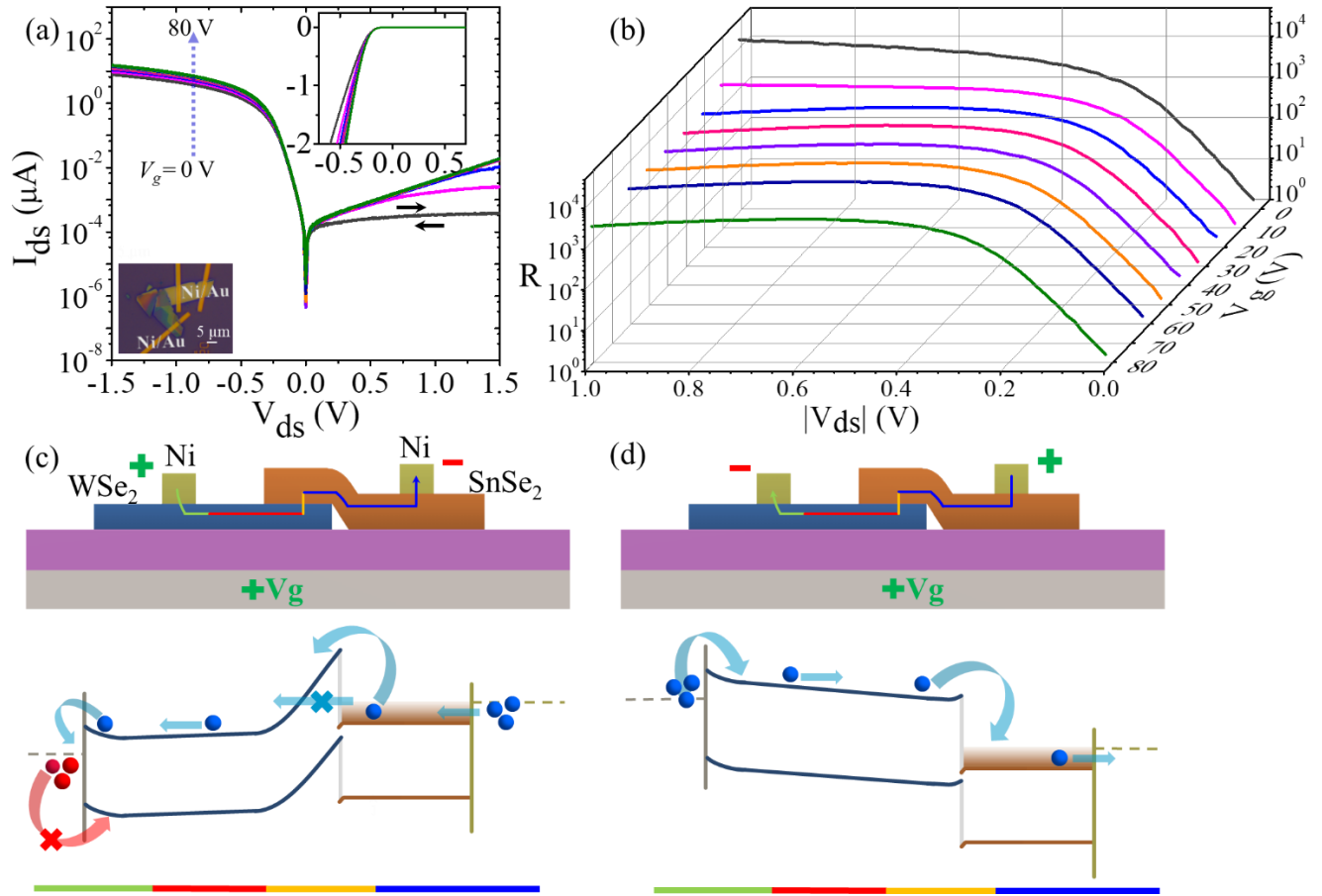

**Figure 6:** (a) $I_{ds}$-$V_{ds}$ curves of device D2 in logarithmic scale at different positive gate voltages from 0 to 80 V varying in steps of 10 V. This device has both of the contacts as Ni/Au. The black arrows indicate the forward and reverse sweep, with negligible hysteresis. Inset, Linear scale plot of $I_{ds}$-$V_{ds}$ showing excellent backward rectification. (b) Reverse rectification ratio ($R$) of D2 as a function of positive gate voltage and drain bias. (c)-(d) Band diagrams under positive gating, with (c) forward and (d) reverse bias, along with schematics of device cross section. The Color bar below each band diagram and mutli-colored arrow in the respective device cross-section map the relevant regions along which transport is considered.

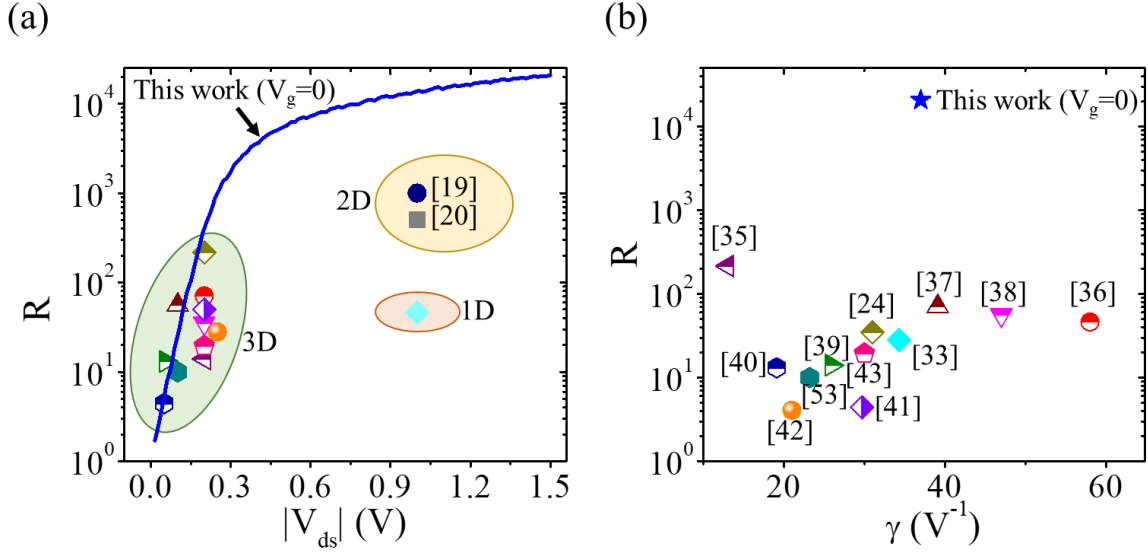

**Figure 7:** (a) Reverse rectification ratio ($R$) of different backward diodes reported in literature plotted (using symbols) as function of applied bias. The data from diodes based on 3D (Si, Ge, III-V), 2D (vdW materials) and 1D (nanotube) materials are clustered separately. The corresponding references for 3D and 1D systems are shown in (b). The solid blue line indicate $R$ from device D2 reported in this work. (b) $R$ versus $\gamma$ benchmarking plot for different reports in literature. The blue star indicates data from device D2.

In Figure 7a, we benchmark the obtained $R$ with backward diode data from literature, by plotting $R$ as a function of applied bias. Our device exhibits more than an order of magnitude higher $R$ compared with the best reported numbers, encompassing backward diodes based on 1D (nanotube), 2D (vdW materials) and 3D (Si, Ge, III-V) systems. Another unique feature of the reported device is that such a large $R$ is maintained even at a very large bias of 1.5 V by suppressing the forward current efficiently, thanks to the large band offset between WSe₂ and SnSe₂. Curvature coefficient ($\gamma$) is another important figure of merit which determines the non-linearity and current sensitivity of a backward diode for detector applications[34,54], and is expressed as

$$\gamma = \frac{d^2 I_{ds} \big/ dV_{ds}^2}{dI_{ds} \big/ dV_{ds}} \bigg|_{V_{ds}=0}$$

We achieve an impressive value of $\gamma$ as 37 V$^{-1}$ for diode D2, which is very close to the theoretical limit of 38.6 V$^{-1}$ at 300 K for transport mechanism controlled by thermionic barrier. In Figure 7b, we benchmark our device with literature data in a $R$ versus $\gamma$ chart.

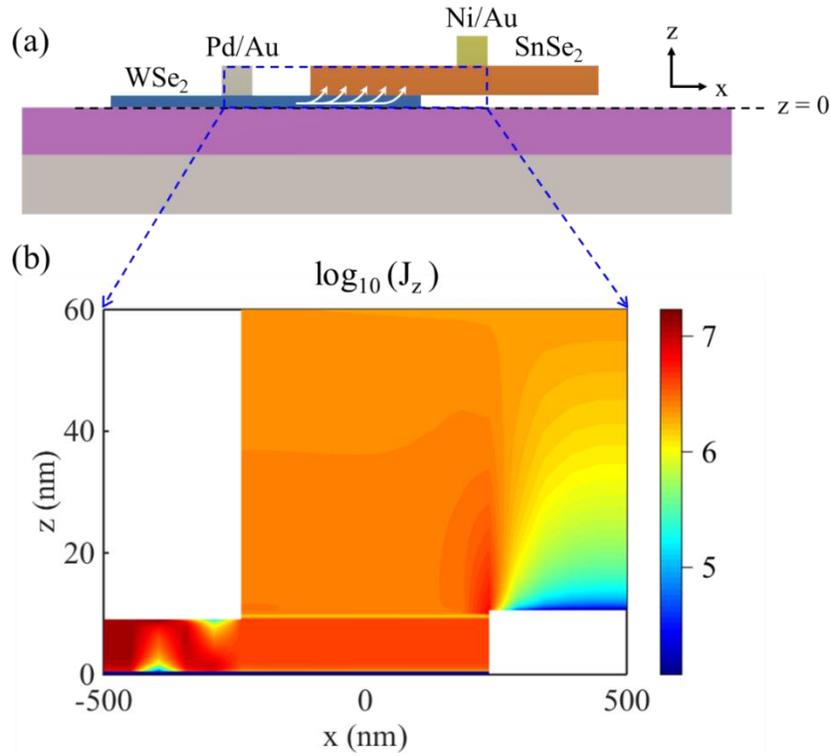

**Figure 8**: (a) Representation of device geometry (highlighted with blue dashed lines in the schematic) used in the simulation of current density distribution. The white arrows highlight the uniform transfer of current in the z direction across the entire interface. (b) Simulated distribution of $z$ component of current density in log scale, suggesting uniform current transfer along the interface.

Finally, we comment on the transfer efficiency of the current from one layer to another in the WSe$_2$/SnSe$_2$ vertical heterojunction. Such a transfer depends on the relative values of the in-plane and out-of-plane conductivity of each material at the interface[55]. We model the current distribution across the entire structure in Figure 8a using the solution of current continuity equation[56]. Conductivity values at different points along the $z$-direction are obtained from the corresponding carrier densities of solved poisson equation. The 2D current density distribution, $J_z(x, z)$, is plotted in Figure 8b, depicting how the current flows across the interface. The current density plot indicates that the transfer of current happens almost uniformly across the entire overlap region of $WSe_2/SnSe_2$. This is in stark contrast with a typical metal-2D semiconductor interface, where the effective contact area is limited by so called *"transfer length"*[44,57], and is only about ~100 nm for metal/TMDs interface[45]. This is understood from the fact that the effective overlap length across which the current is transferred from one material to the other depends on the ratio of the in-plane and the out-of-plane resistivity. Each layer of van der Waals materials offers a resistance to the current flow in its plane which is lower than the resistance offered by the out-of-plane vertical interface across the layers. Consequently, current crowding at such heterointerface is avoided. The entire overlap area is thus important for carrier collection and can be used as a device design parameter to control the total current through the diode.

**Conclusion:**

In conclusion, we demonstrate the highly conducting nature of a gate tunable $WSe_2/SnSe_2$ heterojunction under reverse bias which is appealing for high performance backward diode applications. Drift driven reverse current conduction enables high speed diode operation. By efficiently manipulating the current transport mechanism using gating and device parameter optimization, we achieve a superior backward diode with small reverse threshold ($\sim -0.1$ V), sharp reverse turn on, large reverse rectification ratio ($2.1 \times 10^4$), almost completely suppressed forward current up to a large bias of 1.5 V, and a high curvature coefficient (37 V$^{-1}$). This opens up interesting avenues for high frequency, low noise, radiation hard electronic circuit applications based on vdW backward diode. Further, it is shown that the vertical current transfer efficiency from one layer to the other at the heterojunction interface has a large transfer length, encompassing the entire physical overlap area of the two vdW materials – a result that would be useful for a wide variety of vdW based vertical heterojunctions.

**Experimental Section:**

**Fabrication and characterization of heterojunction:** For PL experiment, we first mechanically exfoliate $SnSe_2$ flake of desired thickness on top of $Si/SiO_2$ substrate. We then exfoliate $WSe_2$ on PDMS sheet and transfer this to a glass slide. Using a micromanipulator, we maneuver the $WSe_2$ flake of interest on PDMS sheet over the $SnSe_2$ flake that is on $Si/SiO_2$ substrate. Then the glass slide with PDMS and substrate are brought close to each other adjusting microscope stage movement in a very slow manner till we observe air release at their interface. At this point, we retract the stage gently to identify a successful formation of $SnSe_2$-bottom/$WSe_2$-top structure. For measuring the electrical characteristics of the back gated heterojunction diode, we swap the $WSe_2$ and $SnSe_2$ layers making a $WSe_2$-bottom/$SnSe_2$-top structure for efficient back gating. Pd/Au (20 nm/50 nm) and Ni/Au (10 nm/50 nm) contacts are patterned by e-beam lithography and deposited with e-beam evaporation technique. All electrical measurements are done using an Agilent B1500 device analyzer keeping the sample chamber pressure less than $10^{-5}$ Torr.

## Acknowledgements

K. M. acknowledges the support of a start-up grant from IISc, Bangalore, the support of grants under Ramanujan Fellowship, Early Career Award, and Nano Mission from the Department of Science and Technology (DST), Government of India.

## References

(1)     Sze, S. .; Ng, K. K. *Physics of Semiconductor Devices*.

(2)     Geim, A. K.; Grigorieva, I. V. Van Der Waals Heterostructures. *Nature* **2013**, *499* (7459), 419–425.

(3)     Li, C.; Zhou, P.; Zhang, D. W. Devices and Applications of van Der Waals Heterostructures. *J. Semicond.* **2017**, *38* (3), 31005.

(4)     Li, M. Y.; Chen, C. H.; Shi, Y.; Li, L. J. Heterostructures Based on Two-Dimensional Layered Materials and Their Potential Applications. *Mater. Today* **2016**, *19* (6), 322–335.

(5)     Pant, A.; Mutlu, Z.; Wickramaratne, D.; Cai, H.; Lake, R. K.; Ozkan, C.; Tongay, S. Fundamentals of Lateral and Vertical Heterojunctions of Atomically Thin Materials. *Nanoscale* **2016**, *8* (7), 3870–3887.

(6)     Shi, Y.; Zhou, W.; Lu, A.-Y.; Fang, W.; Lee, Y.-H.; Hsu, A. L.; Kim, S. M.; Kim, K. K.; Yang, H. Y.; Li, L.-J.; Idrobo, J.-C.; Kong, J. Van Der Waals Epitaxy of $MoS_2$ Layers Using Graphene As Growth Templates. *Nano Lett.* **2012**, *12* (6), 2784–2791.

(7)     Yang, W.; Chen, G.; Shi, Z.; Liu, C.-C.; Zhang, L.; Xie, G.; Cheng, M.; Wang, D.; Yang, R.; Shi, D.; Watanabe, K.; Taniguchi, T.; Yao, Y.; Zhang, Y.; Zhang, G. Epitaxial Growth of Single-Domain Graphene on Hexagonal Boron Nitride. *Nat. Mater.* **2013**, *12* (9), 792–

797.

(8)     Song, X.; Guo, Z.; Zhang, Q.; Zhou, P.; Bao, W.; Zhang, D. W. Progress of Large-Scale Synthesis and Electronic Device Application of Two-Dimensional Transition Metal Dichalcogenides. *Small* **2017**, *13* (35), 1700098.

(9)     Moriya, R.; Yamaguchi, T.; Inoue, Y.; Morikawa, S.; Sata, Y.; Masubuchi, S.; Moriya, R.; Yamaguchi, T.; Inoue, Y.; Morikawa, S.; Sata, Y. Large Current Modulation in Exfoliated-Graphene /$MoS_2$ / Metal Vertical Heterostructures. *Appl. Phys. Lett.* **2014**, *105* (8), 83119.

(10)    Lee, G. H.; Yu, Y. J.; Cui, X.; Petrone, N.; Lee, C. H.; Choi, M. S.; Lee, D. Y.; Lee, C.; Yoo, W. J.; Watanabe, K.; Taniguchi, T.; Nuckolls, C.; Kim, P.; Hone, J. Flexible and Transparent $MoS_2$ Field-Effect Transistors on Hexagonal Boron Nitride-Graphene Heterostructures. *ACS Nano* **2013**, *7* (9), 7931–7936.

(11)    Li, C.; Yan, X.; Bao, W.; Ding, S.; Zhang, D. W.; Zhou, P. Low Sub-Threshold Swing Realization with Contacts of Graphene/h-BN/$MoS_2$ Heterostructures in $MoS_2$ Transistors. *Appl. Phys. Lett.* **2017**, *111*, 193502.

(12)    Li, C.; Yan, X.; Song, X.; Bao, W.; Ding, S. $WSe_2$/$MoS_2$ and $MoTe_2$/$SnSe_2$ van Der Waals Heterostructure Transistors with Different Band Alignment. *Nanotechnology* **2017**, *28*, 415201.

(13)    Roy, K.; Padmanabhan, M.; Goswami, S.; Sai, T. P.; Ramalingam, G.; Raghavan, S.; Ghosh, A. Graphene–$MoS_2$ Hybrid Structures for Multifunctional Photoresponsive Memory Devices. *Nat. Nanotechnol.* **2013**, *8* (11), 826–830.

(14)    Cheng, R.; Li, D.; Zhou, H.; Wang, C.; Yin, A.; Jiang, S.; Liu, Y.; Chen, Y.; Huang, Y.;

Duan, X. Electroluminescence and Photocurrent Generation from Atomically Sharp WSe$_2$/MoS$_2$ Heterojunction P-N Diodes. *Nano Lett.* **2014**, *14* (10), 5590–5597.

(15) Pospischil, A.; Furchi, M. M.; Mueller, T. Solar-Energy Conversion and Light Emission in an Atomic Monolayer P–n Diode. *Nat. Nanotechnol.* **2014**, *9* (4), 257–261.

(16) Britnell, L.; Ribeiro, R. M.; Eckmann, A.; Jalil, R.; Belle, B. D.; Mishchenko, A.; Kim, Y.; Gorbachev, R. V; Georgiou, T.; Morozov, S. V; Grigorenko, A. N.; Geim, A. K.; Casiraghi, C.; Neto, A. H. C.; Novoselov, K. S. Strong Light-Matter Interactions Thin Films. **2013**, *340* (June), 1311–1315.

(17) Ross, J. S.; Klement, P.; Jones, A. M.; Ghimire, N. J.; Yan, J.; Mandrus, D. G.; Taniguchi, T.; Watanabe, K.; Kitamura, K.; Yao, W.; Cobden, D. H.; Xu, X. Electrically Tunable Excitonic Light-Emitting Diodes Based on Monolayer WSe$_2$ P–n Junctions. *Nat. Nanotechnol.* **2014**, *9* (4), 268–272.

(18) Withers, F.; Del Pozo-Zamudio, O.; Mishchenko, A.; Rooney, A. P.; Gholinia, A.; Watanabe, K.; Taniguchi, T.; Haigh, S. J.; Geim, A. K.; Tartakovskii, A. I.; Novoselov, K. S. Light-Emitting Diodes by Band-Structure Engineering in van Der Waals Heterostructures. *Nat. Mater.* **2015**, *14* (3), 301–306.

(19) Roy, T.; Tosun, M.; Cao, X.; Fang, H.; Lien, D.-H.; Zhao, P.; Chen, Y.-Z.; Chueh, Y.-L.; Guo, J.; Javey, A. Dual-Gated MoS$_2$ /WSe$_2$ van Der Waals Tunnel Diodes and Transistors. *ACS Nano* **2015**, *9* (2), 2071–2079.

(20) Liu, X.; Qu, D.; Li, H.-M.; Moon, I.; Ahmed, F.; Kim, C.; Lee, M.; Choi, Y.; Cho, J. H.; Hone, J. C.; Yoo, W. J. Modulation of Quantum Tunneling via a Vertical Two Dimensional Black Phosphorus and Molybdenum Disulfide PN Junction. *ACS Nano* **2017**, *11* (9), 9143–

9150.

(21)     Nourbakhsh, A.; Zubair, A.; Dresselhaus, M. S.; Palacios, T. Transport Properties of a $MoS_2$ /$WSe_2$ Heterojunction Transistor and Its Potential for Application. *Nano Lett.* **2016**, *16* (2), 1359–1366.

(22)     Torrey, H. C.; Whitmer, C. A. *Crystal Rectifiers*; 1948.

(23)     Hopkins, J. B. Microwave Backward Diodes in InAs. *Solid State Electron.* **1970**, *13* (5), 697–705.

(24)     Jin, N.; Yu, R.; Chung, S. Y.; Berger, P. R.; Thompson, P. E.; Fay, P. High Sensitivity Si-Based Backward Diodes for Zero-Biased Square-Law Detection and the Effect of Post-Growth Annealing on Performance. *IEEE Electron Device Lett.* **2005**, *26* (8), 575–578.

(25)     A.B. Bhattacharyya; S.L. Sarnot. Switching Time Analysis of Backward Diodes. In *Proceedings of the IEEE*; 1970; Vol. 58, pp 513–515.

(26)     Burrus, C. A. Backward Diodes for Low-Level Millimeter-Wave Detection. *IEEE Trans. Microw. Theory Tech.* **1963**, *11* (5), 357–362.

(27)     Eng, T.; If, M. LowONoise Properties of Microwave Backward Diodes. In *IRE TRANSACTIONS ON MICROWAVE THEORY AND TECHNIQUES*; 1961; pp 419–425.

(28)     Schlaf, R.; Lang, O.; Pettenkofer, C.; Jaegermann, W. Band Lineup of Layered Semiconductor Heterointerfaces Prepared by van Der Waals Epitaxy: Charge Transfer Correction Term for the Electron Affinity Rule. *J. Appl. Phys.* **1999**, *85* (5), 2732–2753.

(29)     Das, S.; Appenzeller, J. $WSe_2$ Field Effect Transistors with Enhanced Ambipolar Characteristics. *Appl. Phys. Lett.* **2013**, *103* (10), 103501.

(30)  Yan, X.; Liu, C.; Li, C.; Bao, W.; Ding, S.; Zhang, D. W.; Zhou, P. Tunable SnSe$_2$ /WSe$_2$ Heterostructure Tunneling Field Effect Transistor. *Small* **2017**, *13* (34), 1701478.

(31)  Roy, T.; Tosun, M.; Hettick, M.; Ahn, G. H.; Hu, C.; Javey, A. 2D-2D Tunneling Field-Effect Transistors Using WSe$_2$ /SnSe$_2$ Heterostructures. *Appl. Phys. Lett.* **2016**, *108* (8), 83111.

(32)  Li, M. O.; Esseni, D.; Nahas, J. J.; Jena, D.; Xing, H. G. Two-Dimensional Heterojunction Interlayer Tunneling Field Effect Transistors (Thin-TFETs). *IEEE J. Electron Devices Soc.* **2015**, *3* (3), 200–207.

(33)  Liu, C. H.; Wu, C. C.; Zhong, Z. A Fully Tunable Single-Walled Carbon Nanotube Diode. *Nano Lett.* **2011**, *11* (4), 1782–1785.

(34)  Agarwal, S.; Yablonovitch, E. Band-Edge Steepness Obtained From Esaki / Backward Diode Current – Voltage Characteristics. *IEEE Trans. Electron Devices* **2014**, *61* (5), 1488– 1493.

(35)  Okumura, H.; Martin, D.; Malinverni, M.; Grandjean, N.; Okumura, H.; Martin, D.; Malinverni, M.; Grandjean, N. Backward Diodes Using Heavily Mg-Doped GaN Growth by Ammonia Molecular-Beam Epitaxy. *Appl. Phys. Lett.* **2016**, *108*, 72102.

(36)  Rahman, S. M.; Jiang, Z.; Fay, P.; Liu, L.; Rahman, S. M.; Jiang, Z.; Fay, P.; Liu, L. Integration and Fabrication of High-Performance Sb-Based Heterostructure Backward Diodes with Submicron-Scale Airbridges for Terahertz Detection. *J. Vac. Sci. Technol. B* **2016**, *43* (4), 41220.

(37)  Meyers, R. G.; Fay, P.; Schulman, J. N.; Thomas, S.; Chow, D. H.; Zinck, J.; Boegeman,

Y. K.; Deelman, P. Bias and Temperature Dependence of Sb-Based Heterostructure Millimeter-Wave Detectors with Improved Sensitivity. *IEEE Electron Device Lett.* **2004**, *25* (1), 4–6.

(38)    Zhang, Z.; Rajavel, R.; Deelman, P.; Fay, P. Sub-Micron Area Heterojunction Backward Diode Millimeter-Wave Detectors with 0.18 pW/Hz1/2 Noise Equivalent Power. *IEEE Microw. Wirel. Components Lett.* **2011**, *21* (5), 267–269.

(39)    Fay, P.; Schulman, J. N.; Thomas, S.; Chow, D. H.; Boegeman, Y. K.; Holabird, K. S. High-Performance Antimonide-Based Heterostructure Backward Diodes for Millimeter-Wave Detection. *IEEE Electron Device Lett.* **2002**, *23* (10), 585–587.

(40)    Schulman, J. N.; Chow, D. H. Sb-Heterostructure Interband Backward Diodes. *IEEE Electron Device Lett.* **2000**, *21* (7), 353–355.

(41)    Schulman, J. N.; Chow, D. H.; Jang, D. M. InGaAs Zero Bias Backward Diodes for Millimeter Wave Direct Detection. *IEEE Electron Device Lett.* **2001**, *22* (5), 200–202.

(42)    Simon, J.; Zhang, Z.; Goodman, K.; Xing, H.; Kosel, T.; Fay, P.; Jena, D. Polarization-Induced Zener Tunnel Junctions in Wide-Band-Gap Heterostructures. *Phys. Rev. Lett.* **2009**, *103* (2), 26801.

(43)    Yu, T.; Teherani, J. T.; Antoniadis, D. A.; Hoyt, J. L. In0.53 Ga0.47 As GaAs0.5 Sb0.5 Quantum-Well Tunnel-FETs with Tunable Backward Diode Characteristics. *IEEE Electron Device Lett.* **2013**, *34* (12), 1503–1505.

(44)    Somvanshi, D.; Kallatt, S.; Venkatesh, C.; Nair, S.; Gupta, G.; Anthony, J. K.; Karmakar, D.; Majumdar, K. Nature of Carrier Injection in metal/2D Semiconductor Interface and Its Implications to the Limits of Contact Resistance. *Phys. Rev. B-solid state* **2017**, *96*, 205423.

(45)  English, C. D.; Shine, G.; Dorgan, V. E.; Saraswat, K. C.; Pop, E. Improved Contacts to MoS₂ Transistors by Ultra-High Vacuum Metal Deposition. *Nano Lett.* **2016**, *16* (6), 3824–3830.

(46)  Peng, B.; Yu, G.; Liu, X.; Liu, B.; Liang, X.; Bi, L.; Deng, L.; Sum, T. C.; Loh, K. P. Ultrafast Charge Transfer in MoS ₂ /WSe ₂ P–n Heterojunction. *2D Mater.* **2016**, *3* (2), 25020.

(47)  Ceballos, F.; Bellus, M. Z.; Chiu, H. Y.; Zhao, H. Ultrafast Charge Separation and Indirect Exciton Formation in a MoS₂-MoSe₂ van Der Waals Heterostructure. *ACS Nano* **2014**, *8* (12), 12717–12724.

(48)  Zhang, X.; Tan, Q.-H.; Wu, J.-B.; Shi, W.; Tan, P.-H. Review on the Raman Spectroscopy of Different Types of Layered Materials. *Nanoscale* **2016**, *8* (12), 6435–6450.

(49)  Aretouli, K. E.; Tsoutsou, D.; Tsipas, P.; Marquez-Velasco, J.; Aminalragia Giamini, S.; Kelaidis, N.; Psycharis, V.; Dimoulas, A. Epitaxial 2D SnSe₂/ 2D WSe₂ van Der Waals Heterostructures. *ACS Appl. Mater. Interfaces* **2016**, *8* (35), 23222–23229.

(50)  Krishna, M.; Kallatt, S.; Majumdar, K. Substrate Effects in High Gain , Low Operating Voltage SnSe₂ Photoconductor. *Nanotechnology* **2018**, *29*, 35205.

(51)  Robertson, J. Electronic Structure of SnS₂ , SnSe₂ , CdI₂ and PbI₂. *J. Phys. C Solid State Phys.* **1979**, *12* (22), 4753–4766.

(52)  Coehoorn, R.; Haas, C.; Dijkstra, J.; Flipse, C. J. F.; De Groot, R. A.; Wold, A. Electronic Structure of MoSe₂, MoS₂, and WSe₂. I. Band-Structure Calculations and Photoelectron

Spectroscopy. *Phys. Rev. B* **1987**, *35* (12), 6195–6202.

(53)    S.-Y. Park; Yu, R.; Chung, S.-Y.; Berger, P. R.; Thompson P.E; Fay, P. Sensitivity of Si-Based Zero-Bias Backward Diodes for Microwave Detection. *Electron. Lett.* **2007**, *43* (5).

(54)    Karlovsky. J. The Curvature Coefficient of Germanium Tunnel and Backward Diodes. *Solid State Electron.* **1967**, *10*, 1109–1111.

(55)    D K Schroder. *Semiconductor Material and Device Characterization*; 2006.

(56)    Majumdar, K.; Vivekanand, S.; Huffman, C.; Matthews, K.; Ngai, T.; Chen, C. H.; Baek, R. H.; Loh, W. Y.; Rodgers, M.; Stamper, H.; Gausepohl, S.; Kang, C. Y.; Hobbs, C.; Kirsch, P. D. STLM: A Sidewall TLM Structure for Accurate Extraction of Ultralow Specific Contact Resistivity. *IEEE Electron Device Lett.* **2013**, *34* (9), 1082–1084.

(57)    Xia, F.; Perebeinos, V.; Lin, Y.; Wu, Y.; Avouris, P. The Origins and Limits of Metal–graphene Junction Resistance. *Nat. Nanotechnol.* **2011**, *6* (3), 179–184.

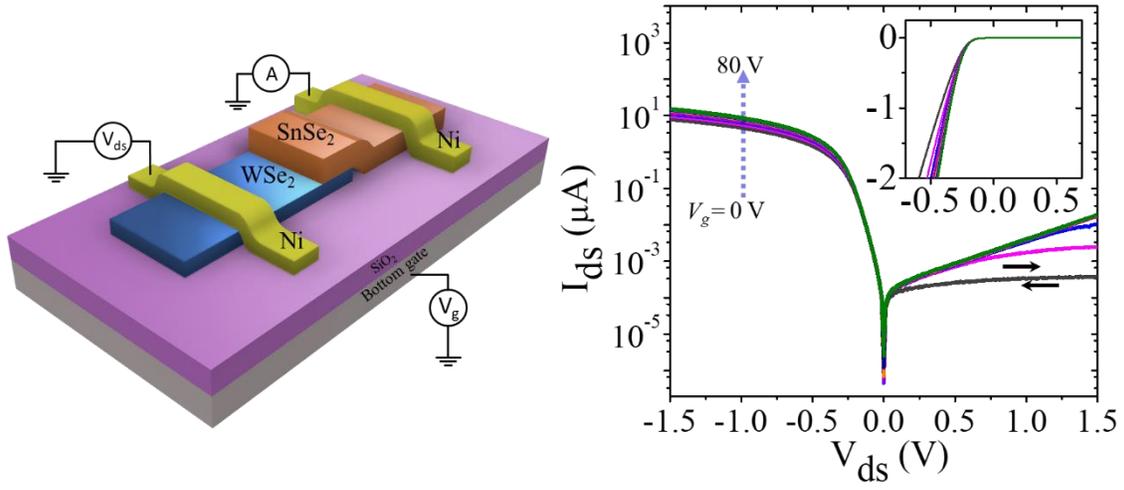

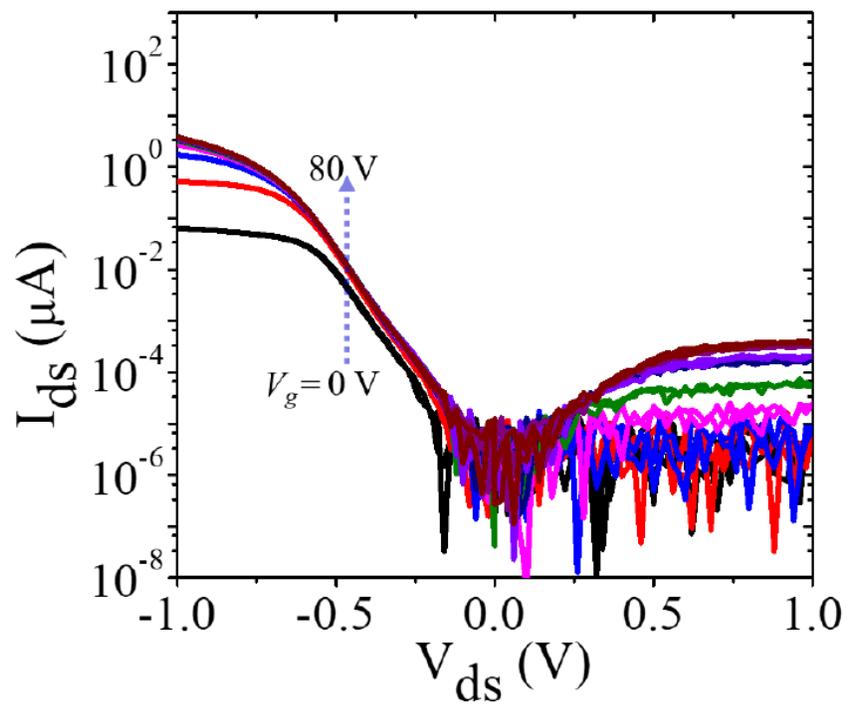

Figure S1: Characteristics of a Ni contacted WSe₂/SnSe₂ backward diode, obtained from a different run, showing large reverse current and suppressed forward current, giving rise to large reverse rectification ratio.